\documentclass[conference]{IEEEtran}
\IEEEoverridecommandlockouts

\usepackage[utf8]{inputenc}
\usepackage{booktabs}
\usepackage{graphicx}
\usepackage{subfig}
\usepackage{xspace}
\usepackage{siunitx}
\usepackage{listings}

\usepackage{amsmath,amssymb,amsfonts}
\usepackage{algorithmic}
\usepackage{graphicx}
\usepackage{textcomp}
\usepackage{xcolor}
\usepackage{multirow}
\usepackage{graphicx}
\PassOptionsToPackage{hyphens}{url}\usepackage{hyperref}
\usepackage[utf8]{inputenc}
\usepackage{booktabs}
\usepackage{subfig}
\usepackage{xspace}
\usepackage{siunitx}

\makeatother
\def\BibTeX{{\rm B\kern-.05em{\sc i\kern-.025em b}\kern-.08em
    T\kern-.1667em\lower.7ex\hbox{E}\kern-.125emX}}

\usepackage{amssymb}
\usepackage{pifont}
\newcommand{\pb}[1]{\vspace{0.75ex}\noindent{\bf \em #1}\hspace*{.3em}}

\def\eg{\emph{e.g. }\xspace}
\def\etc{\emph{etc. }\xspace}
\def\ie{\emph{i.e.,}\xspace}
\def\etal{\emph{et al.}\xspace}
\def\vs{\emph{vs.}\xspace}

\newcommand{\one}{({\em i}\/)}
\newcommand{\two}{({\em ii}\/)}
\newcommand{\three}{({\em iii}\/)}
\newcommand{\four}{({\em iv}\/)}

\newcounter{RZNumberOfComments}
\stepcounter{RZNumberOfComments}

\usepackage[markup=underlined]{changes}
\usepackage{todonotes}

\definechangesauthor[color=red]{Koosha}
\definechangesauthor[color=blue]{Reza}
\definechangesauthor[color=magenta]{Gareth}
\definechangesauthor[color=magenta]{Kiran}
\definechangesauthor[color=magenta]{Zafar}

\setcommentmarkup{\todo[color={authorcolor!20},size=\scriptsize]{#3: #1}}

\begin{document}

\title{Characterising and Detecting Sponsored Influencer Posts on Instagram}

\vspace{-0.8cm}
\author{\IEEEauthorblockN{Koosha Zarei\IEEEauthorrefmark{1}, Damilola Ibosiola\IEEEauthorrefmark{2}, Reza Farahbakhsh\IEEEauthorrefmark{1}, Zafar	Gilani\IEEEauthorrefmark{2}, Kiran	Garimella\IEEEauthorrefmark{3},  No\"{e}l Crespi\IEEEauthorrefmark{1}, Gareth Tyson\IEEEauthorrefmark{2}}
\IEEEauthorblockA{\IEEEauthorrefmark{1}\textit{Institut Polytechnique de Paris, T\'el\'ecom SudParis} \textit{France}.
\{koosha.zarei, reza.farahbakhsh, noel.crespi\}@telecom-sudparis.eu}
\IEEEauthorrefmark{2}\textit{Queen Mary University of London, United Kingdom}. \{d.i.ibosiola, z.gilani, gareth.tyson\}@qmul.ac.uk \\
\IEEEauthorrefmark{3}\textit{Massachusetts Institute of Technology}. garimell@mit.edu
}

\maketitle

\begin{abstract}  

Recent years have seen a new form of advertisement campaigns emerge: those involving so-called social media \emph{influencers}. These influencers accept money in return for promoting products via their social media feeds.
Although this constitutes a new and interesting form of marketing, it also raises many questions, particularly related to transparency and regulation. For example, it can sometimes be unclear which accounts are officially influencers, or what even constitutes an influencer/advert.
This is important in order to establish the integrity of influencers and to ensure compliance with advertisement regulation.
We gather a large-scale Instagram dataset covering thousands of accounts advertising products, and create a categorisation based on the number of users they reach. 
We then provide a detailed analysis of the types of products being advertised by these accounts, their potential reach, and the engagement they receive from their followers.
Based on our findings, we train machine learning models to distinguish sponsored content from non-sponsored, and identify cases where people are generating sponsored posts without officially labelling them. Our findings provide a first step towards understanding the under-studied space of online influencers that could be useful for researchers, marketers and policymakers.

\end{abstract}

\begin{IEEEkeywords}
Influencers, Instagram, Social Media, User Behaviour Analysis, Sponsored Content
\end{IEEEkeywords}

\section{Introduction}\label{introduction_section}
Social media has become an active part of billions of people's lives. Initially intended as a method to interact amongst friends, it has since become a platform used heavily for marketing. This is dominated by players such as Instagram who allow third parties to micro-target adverts at users matching certain criteria. This newfound ability has turned social media into a multi-billion dollar industry. 
However, in recent years a new breed of social advertisement has emerged. Driven by the rise of social media celebrities, so-called \emph{influencers}, we have witnessed various companies marketing directly through celebrity endorsements (rather than via official advertisement channels on the platform).
For example, this may involve a celebrity posting an image of them wearing a particular item of clothing or driving a particular car. Whereas endorsement by celebrities is nothing new, social platforms have enabled this to take place on an industrial scale. The proliferation of these activities, via the use of Influencer Markets,\footnote{\url{https://influencermarketinghub.com/influencer-marketing-2019-benchmark-report/}} has also dramatically increased the scale and accessibility of these types of promotion. 

A particularly interesting trend is the rise of \emph{nano} influencers \cite{CASALO2018}. These users may only have a small number of followers, but often have reach into highly targeted audiences. This has radically increased the number of accounts actively promoting products, thereby making the industry challenging to regulate. This is because such social media posts are highly opaque, often making it impossible to identify the difference between paid promotions, and genuine personal endorsements. This distinction is important, as personal endorsements do not fall under the same stringent regulatory framework as paid sponsorship.
To address this, standards bodies such as the UK's Advertising Standards Authority (ASA) have created a set of rules asking that such post are explicitly tagged with relevant hashtags (\eg \#ad)~\cite{asa_guidelines}. 
Similarly platforms such as Instagram, have allowed such information to be formally embedded within posts via their user interface. To date, however, the research community has limited knowledge of how these practices are employed, or even how influencer marketing takes place more generally.

This paper strives to fill this gap, and makes two contributions. \emph{First}, we present a broad characterisation of influencers on Instagram, and formally quantify their behavioural attributes. Although inherently of value, we also argue that this has a number of implications for emerging regulation in this sector. 
To this end, we collect a large-scale Instagram dataset consisting of 12k influencers (\S\ref{sec:dataset}). 
We gather both Instagram posts and stories\footnote{Stories are time limited posts that automatically delete after 24 hours.} from  users who attach advert-related hashtags to their material.



With this dataset, we proceed to explore influencer characteristics (\S\ref{sec:char}).
We see a range of promotion activity taking place, ranging from ``mega'' influencers with over 1 million followers to ``nano'' influencers with as little as 500 followers. We find a range of products being produced with health \& beauty taking the top spot, followed by services, clothing and food featuring prominently.
Naturally, top mega influencers garner the greatest attention, as measured via likes and comments. Yet we find that small ``nano'' influencers sustain attention more effectively. For example, we find that a far larger fraction of nano influencers' followers engage with posts, thus forming a denser social network (whereas, in absolute terms, higher popularity celebrities gain more likes). This suggests that these different types of influences offer marketers very different styles of promotion. We also observe cases where users are creating sponsored posts, yet failing to attach appropriate labels to inform audiences that they are receiving payment. This not only has the potential to harm vulnerable users (\eg children) but also raises legal questions.  

This drives us to make our \emph{second} contribution (\S\ref{sec:predicting}), where we strive to build automated tools to detect sponsored posts that fail to explicitly tag themselves. 
We propose a Contextual LSTM Neural Network classifier which considers the text content and other metadata to identify whether a post is sponsored or not. We obtain an accuracy of 89\%, and we identify a set of prominent predictive features, \eg{}use of language, the inclusion of URLs and commentary.
Our findings have important implications for social media companies, marketers, researchers, and regulators wishing to better understand the behaviour of influencers.

\section{Background \& Related Work}
The concept of social influence is a well studied topic. It has long been known that an elite of \emph{influencers} can have a disproportionate impact on social structures~\cite{rogers2010diffusion, SINGH201987}.
This has subsequently led to the rise of \emph{influencers}, who monetise their influence via sponsorship payments. 
The premise is that, via electronic word of mouth, brands can be promoted through influential endorsements, akin to traditional sponsorship deals~\cite{hennig2004electronic}.
More formally, sponsored influencers are ``third party individuals who significantly shape the customer’s purchasing decision, but may never be accountable for it''~\cite{brown2008influencer}. 
There have been a number of recent studies looking at influencer marketing, primarily within the business research community. 
These have followed a number of methodologies, including interviews with marketers~\cite{biaudet2017influencer,ewers2017sponsored} and customers~\cite{vaibhavi_nandagiri_2018_1207039}, as well as qualitative studies of specific influencers~\cite{glucksman2017rise}.
There have been further preliminary studies of the perception of influencers that, for example, found people to view influencers as smart and ambitious~\cite{freberg2011social, sammis2015influencer}.

There have also been a small set of studies that rely on computational datasets. Lahuerta \etal~\cite{lahuerta2016looking} studied the role of marketing influencers in the Japanese automotive sector. They collected 30K tweets based on relevant keywords. They found, amongst other things, that the sentiment and length of influencer tweets has a positive effect on popularity. 
Although highly relevant to our work, it is unclear how these trends generalise across other brands and large-scale influencer networks. Closest to our work is~\cite{kim2017social}, which explored the follower graphs of influencers on Instagram. 
They targeted 218 specific social influencers (with 8.9M followers) as registered via Popular Pays, a platform used to connecting brands to influencers. These were then compared against 948 Instagram users with fewer than 10K followers. 
They studied the social graph properties of these users, to find that influencers have large numbers of followers, and even tend to interconnect amongst themselves. They also found that users often follow multiple influencers. 
Although highly applicable, we argue that these targeted studies likely only capture a small slice of the overall influencer market. They have a particular bias towards certain types of highly popular influencers, and ignore a potentially larger population of micro influencers. To the best of our knowledge, this is the first large-scale study of influencer activities. 


\section{Data Collection Methodology}
\label{sec:dataset}

We start by explaining our data collection methodology. 
Our data collection covers both Instagram posts and the recently introduced stories feature. 

\subsection{Data Collection}
\label{instagram_crawler}

Our data collection activities took place over four phases:

\pb{Phase 1: Hashtags.}
It is first necessary to obtain a large list of influencer accounts. One approach would be to manually curate this set, however, this would limit us to a small set of influencers, largely dominated by well known celebrities who are easy to identify. Hence, we compile the list by crawling all posts attached to a set of influencer-related hashtags. 
We turn to the UK's Advertising Standards Agency~\cite{asa_guidelines}, which states that influencers should use the \#ad, \#advert or \#sponsored hashtags in \emph{any} posts that have been paid for. We expand this list with \#advertising, \#giveaway, \#spon and \#sponsor ~\cite{FTC:online}.

\pb{Phase 2: Post \& Stories Collection.}
We then use the official Instagram API \cite{InstagramAPI_online} to gather all posts and stories that include any of the above hashtags. Note that stories are similar to normal posts, yet they are automatically deleted after 24 hours (akin to Snapchat posts). Hashtag Engine is used \cite{InstagramAPIHashtag:online} with a maximum of 30 unique hashtags.
This API returns public photos and videos that have been tagged with specific hashtags.
Our crawl for posts and stories ran between Sep 2018 and April 2019.
This process identifies 12K accounts that have posted using the previously mentioned hashtag.

\pb{Phase 3: Account Collection.}
Although the above yields a substantial body of posts and stories, we are primarily interested in gathering data on a per-influencer basis. 
Hence, we next extract all accounts identified from the Phase 2 dataset and begin dedicated monitoring for all posts and stories generated by those users. (\ie influencers).
This covers all posts, reactions and stories from those accounts from July 2019 to August 2019. In this step, we collect 19.7K posts, 63K stories, 3.1M comments, and 27M likes (generated by the 12K user accounts from Phase 2). Note that they contain a mix of both sponsored (16\%) and non-sponsored (84\%) entities.
For each post, we collect the image, comments, likes and public profile information of the user, as well as any other users who reacted to the post.
For each story, we collect the equivalent information, although we cannot collect likes (as these are not available in stories).
Each sponsored post is also tagged with the product being advertised, and the category of advertiser. In total, we have 35K posts, 99K stories, 3.1M comments, and 27M likes generated by 12K users.





\pb{Phase 4: Categorization.}\label{subsection_detect_sponsored}
Once we have collected the posts and stories, it is necessary to tag explicitly which are considered sponsored. We take a simple approach. If a post is tagged with one of the above hashtags, we assume that it is sponsored. In the case of Instagram stories, there is explicit metadata which tells us if it is sponsored (the Paid Partnership tag). Hence, for stories we rely on this metadata item (rather than hashtags). Note that this excludes posts that are sponsored, yet the user does not add the appropriate hashtag. We revisit this issue in \S\ref{sec:predicting}.


\pb{Ethics.} In line with Instagram policies as well as user privacy, we only gather publicly available data that is obtainable from Instagram. Whereas we do analyse the content of influencer posts, we do not inspect the content of comments submitted by non-influencer users.



\subsection{Data Validation}
A natural risk is that a subset of the posts containing the curated hashtags may be generated by users who are not influencers. 
Although it is impossible to entirely discount this at scale, we further perform manual annotation to validate the general correctness of our data. To validate our dataset, we manually looked at the profiles of the influencers to verify if they were really promoting sponsored content. 

All users with above 10K followers are checked, confirming that they were all correctly tagged as posting sponsored content.
We further check 25\% (2K) of all influencers with under 10K followers. We find that we the above approach yields 97.6\% accuracy: just 48 accounts were incorrectly classified as influencers. 
Note that the above only checks if a user account has one more truly sponsored posts. To provide further confidence we randomly select 500 influencers and check \emph{all} of their posts. Around 80\% of sponsored-post are correctly classified as the sponsored content (based on the hashtags previously mentioned). We filter any incorrectly identified influencers.


\section{Characterizing Influencers}
\label{sec:char}

We begin by exploring influencers' follower counts and engagement levels (comments and likes), before profiling the types of products promoted.



\begin{figure}[htbp]
\vspace{-0.4cm}
  \subfloat[All dataset]{\includegraphics[width=0.49\linewidth, , height=0.215\textwidth]{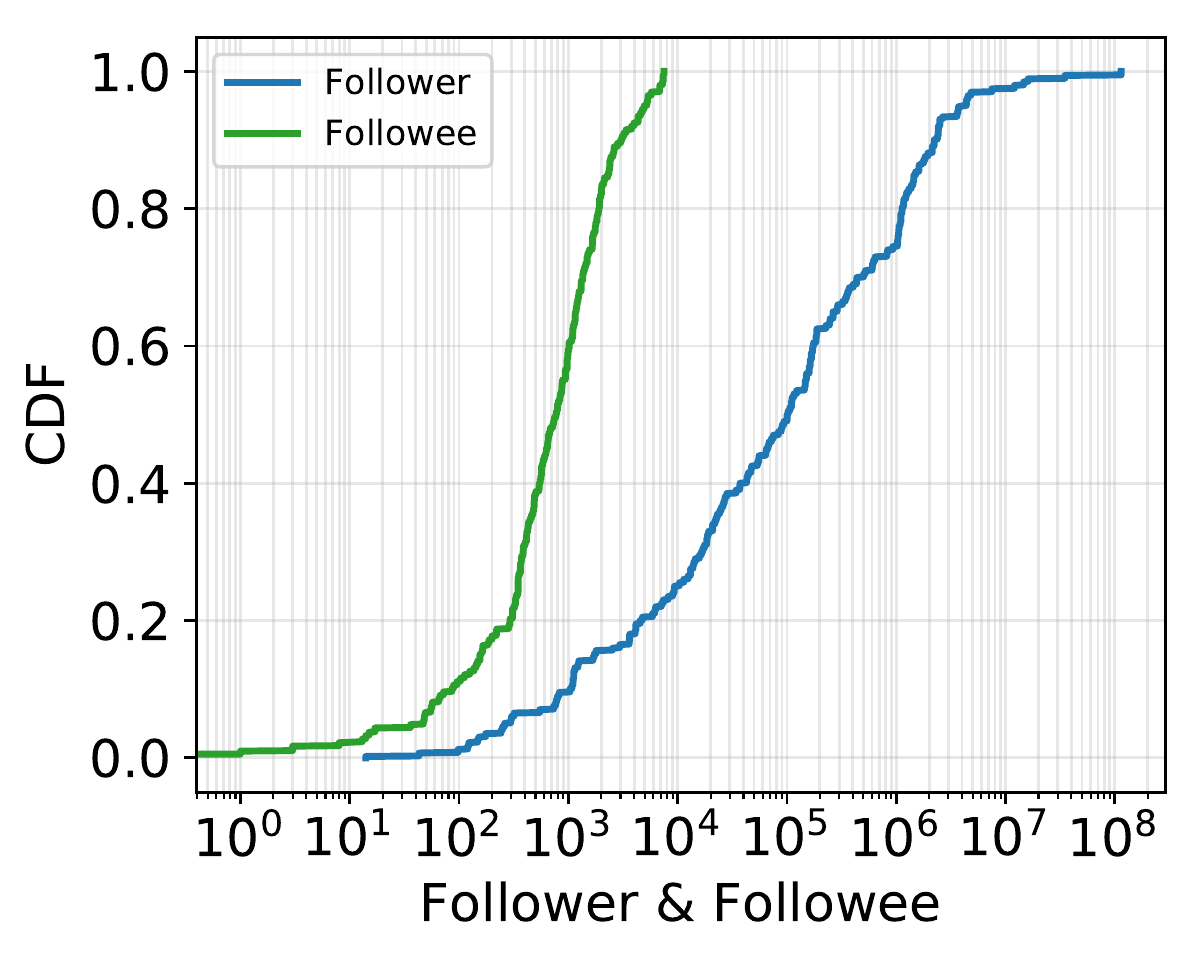}}\hfill
  \subfloat[Follower per account]{\includegraphics[width=0.49\linewidth, height=0.215\textwidth]{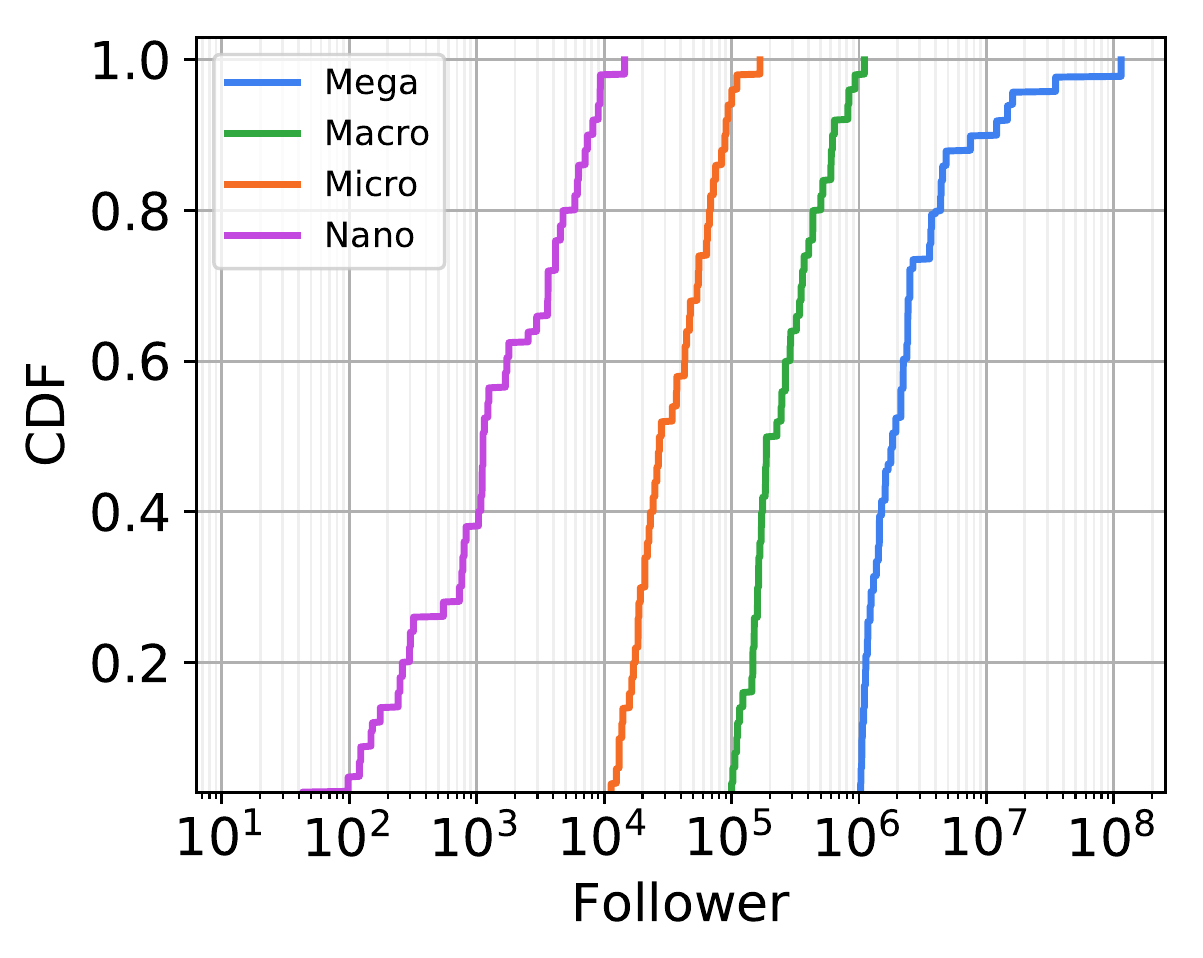}}%
 \vspace{-0.2cm}
\caption{Follower counts: (a) CDF of followers and followees of all the influencers in our data.  (b) CDF of number of followers per account separated in groups.}

\label{fig_instagram_general_dist}
\vspace{-0.3cm}
\end{figure}

\subsection{How popular are influencers?}
\label{how_popular_are_influencers}

We take inspiration from past work~\cite{cha2010measuring}, and begin our analysis of influence by looking at follower counts. This is a natural indicator of influence as it captures the upper-bound of people to whom posts are directly pushed. Figure \ref{fig_instagram_general_dist}(a) presents the cumulative distribution function (CDF) of the follower and followee counts of the influencers in our dataset. 

Unsurprisingly, we see a sizeable fraction of extremely popular accounts. 35\% of users have over 100K followers, and 17\% posses over 1M. These conform to the common interpretation of influencers. More surprising, however, is the presence of a long tail: 37\% of accounts have fewer than 10K influencers, with 15.5\% even having below 1K. 
At first, we suspected that this may be caused by miscellaneous use of the advert-related hashtags. However, upon manual inspection, we confirm that these are indeed influencers. For instance, @lesya\_9\_9 with 500 followers has promoted over 10 times more regularly than @blogging\_with\_tiffany who has fewer than 100 followers. This reveals a growing set of small-scale influencers who promote products, despite their low follower counts. In other words, \emph{influencers are not just celebrities}: they appear to encompass a morass of different account types. 
For context, we can contrast these results with the followee count (Figure \ref{fig_instagram_general_dist}(a)) which are, on average, far lower than follower numbers. Whereas the median follower count is 23.7K, it is just 770 for the number of followees.




\begin{table}[htbp]
\vspace{-0.2cm}
\centering
\caption{Influencer Profile Characteristics}
\vspace{-0.2cm}
\label{table_advertiser_instagram}
\resizebox{\columnwidth}{!}{%
\begin{tabular}{@{}llllll@{}}
\toprule
\multicolumn{1}{c}{\textbf{\begin{tabular}[c]{@{}c@{}}Influencer\\ Category\end{tabular}}} & 
\multicolumn{1}{c}{\textbf{\begin{tabular}[c]{@{}c@{}}\#Followers\\ Category\end{tabular}}} &
\multicolumn{1}{c}{\textbf{\begin{tabular}[c]{@{}c@{}}Avg. \\ follower\end{tabular}}} & \multicolumn{1}{c}{\textbf{\begin{tabular}[c]{@{}c@{}}Avg. \\  followee\end{tabular}}} & \multicolumn{1}{c}{\textbf{\begin{tabular}[c]{@{}c@{}}Avg. \\ mediacount\end{tabular}}} & \multicolumn{1}{c}{\textbf{\begin{tabular}[c]{@{}c@{}}\% of\\ verified\end{tabular}}}   \\ \midrule
\textbf{Mega} & $\geq1M$  & 5.8m & 845 & 9.1k & 82\%  \\
\textbf{Macro} & $<1M$ \& $\geq100K$ & 257k & 1.3k & 3.1k  & 22\% \\
\textbf{Micro} & $<100k$ \& $>10K$ & 32k & 1.9k & 1.8k  & 4\%  \\
\textbf{Nano} & $<10K$ & 1.5k & 0.9k & 597  & 0.5\%  \\ 
\bottomrule
\multicolumn{5}{ c }
{The key profile characteristics of influencers (full timeline).}
\end{tabular}}
\vspace{-0.4cm}
\end{table}


Based on the above findings, we categorise influencers into 4 distinct categories based on their reach (\# followers). This taxonomy underpins our subsequent analysis. We term these nano, micro, macro and mega influencers. 
Table \ref{table_advertiser_instagram} presents a summary of these groups.
We note that 80\% of mega influencers are verified by Instagram, but in contrast, under 5\% of nano and micro accounts have a blue verified icon.\footnote{Users on Instagram can get verified badge as low as 500 followers. However, that account must represent a well-known, highly searched for person, brand or entity \cite{InstagramRqe:online}}
Figure~\ref{fig_instagram_general_dist}(b) presents the CDF of follower counts per-account, broken into these four groups. Naturally, the distributions reflect the split with nano influencers having the fewest followers. For context, a few examples of influencers (top three in terms of followers) from the each categories is shown in Table \ref{dataset_top_influencers}. As previously discussed, primarily the more ``popular'' influencers have verified accounts. These users tend to also have more posts on average.

\begin{table}[htbp]
\centering
\caption{Examples of influencers}
\label{dataset_top_influencers}
\resizebox{\columnwidth}{!}{%
\begin{tabular}{l|llrrrll}
\multicolumn{1}{c|}{\textit{\textbf{Category}}} & \multicolumn{1}{c}{\textit{\textbf{Username}}} & \multicolumn{1}{c}{\textit{\textbf{Section}}} & \multicolumn{1}{c}{\textit{\textbf{\#post}}} & \multicolumn{1}{c}{\textit{\textbf{\#follower}}} & \multicolumn{1}{c}{\textit{\textbf{\#followee}}} & \multicolumn{1}{c}{\textit{\textbf{\#verified}}} & \multicolumn{1}{c}{\textit{\textbf{\#url}}} \\ \hline
\multirow{3}{*}{\textit{\textbf{Mega}}} & @kendalljenner & Fashion & 3K & 116M & 203 & \checkmark & - \\
 & @vanessahudgens & Fashion & 3.2K & 36M & 1.1K & \checkmark & - \\
 & @brentrivera & Lifestyle & 1.8K & 16M & 395 & \checkmark & Youtube \\ \hline
\multirow{3}{*}{\textit{\textbf{Macro}}} & @tonyamichelle26 & Lifestyle & 5.3K & 937K & 2.1K & - & Business Page \\
 & @alice\_gao & Design & 4.1K & 910K & 500 & \checkmark & Business Page \\
 & @lilleejean & Beauty & 700 & 950K & 650 & \checkmark & Youtube \\ \hline
\multirow{3}{*}{\textit{\textbf{Micro}}} & @charisseo\_ & Fashion & 1.7K & 98K & 600 & - & Email \\
 & @morgbullard & Lifestyle & 1.9K & 97.5K & 1.3K & - & Business Page \\
 & @ginascrocca & Fashion & 246 & 96K & 1K & - & Business Page \\ \hline
\multirow{3}{*}{\textit{\textbf{Nano}}} & @jaimesays & Travel & 600 & 9K & 1.9K & - & Business Page \\
 & @aberhalloooo & Food & 1.1K & 9K & 7K & - & Email \\
 & @lawrence.carlyFollow & Dance & 393 & 8.1K & 1.2K & - & Business Page
\end{tabular}
}
\vspace{-0.3cm}
\end{table}

\subsection{Do followers engage with influencers?}
\label{nature_of_engagement}

Another way to measure ``influence'' is to inspect engagement levels on a users' posts, \eg comments, likes and mentions. Previous studies~\cite{cha2010measuring} have argued that these levels can be a better proxy for influencer than simply inspecting follower counts. In this section, we directly contrast engagement levels for posts that are sponsored \vs not sponsored.

\begin{figure}[htbp]
\vspace{-0.4cm}
  \subfloat[Absolute]{\includegraphics[width=0.49\linewidth, , height=0.225\textwidth]{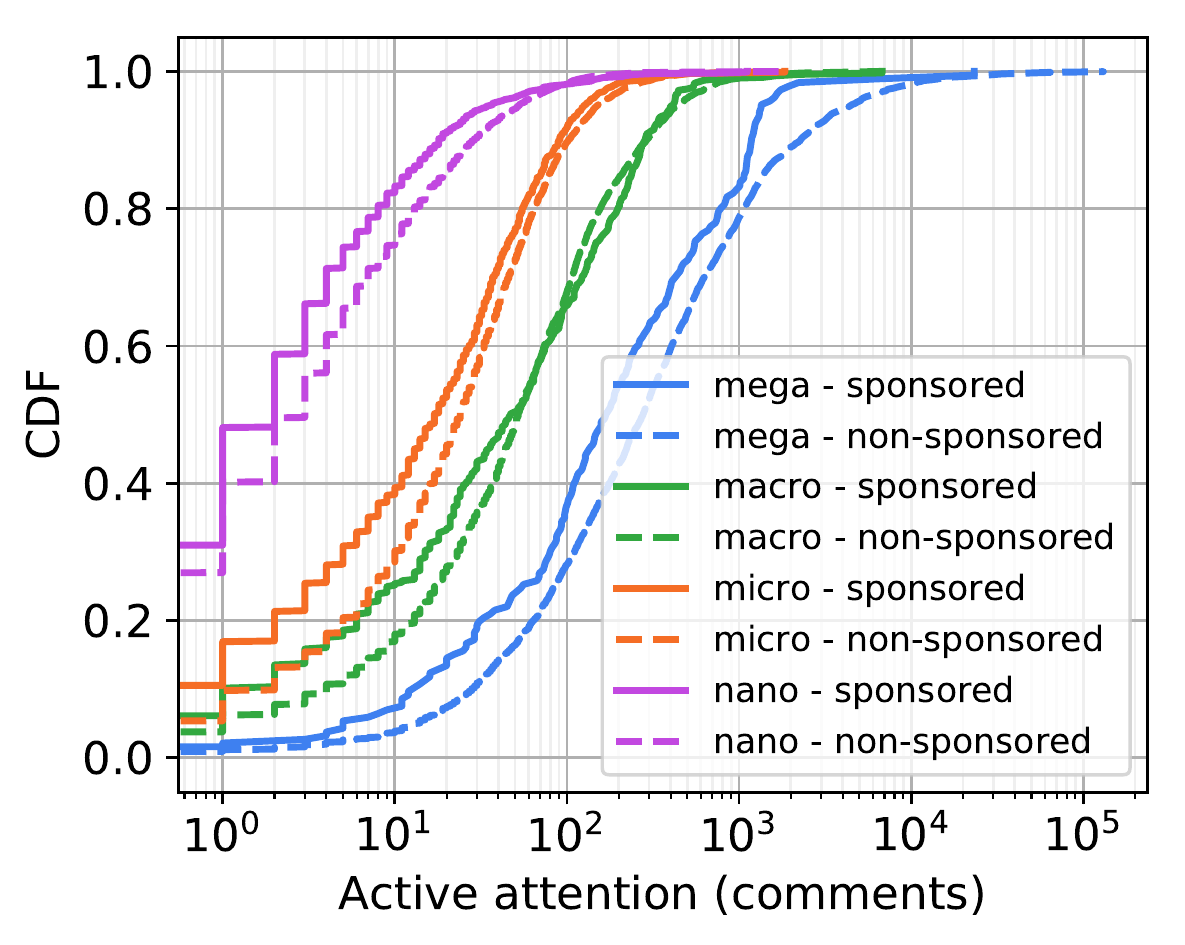}}\hfill
  \subfloat[Normalized by followers]{\includegraphics[width=0.49\linewidth, height=0.225\textwidth]{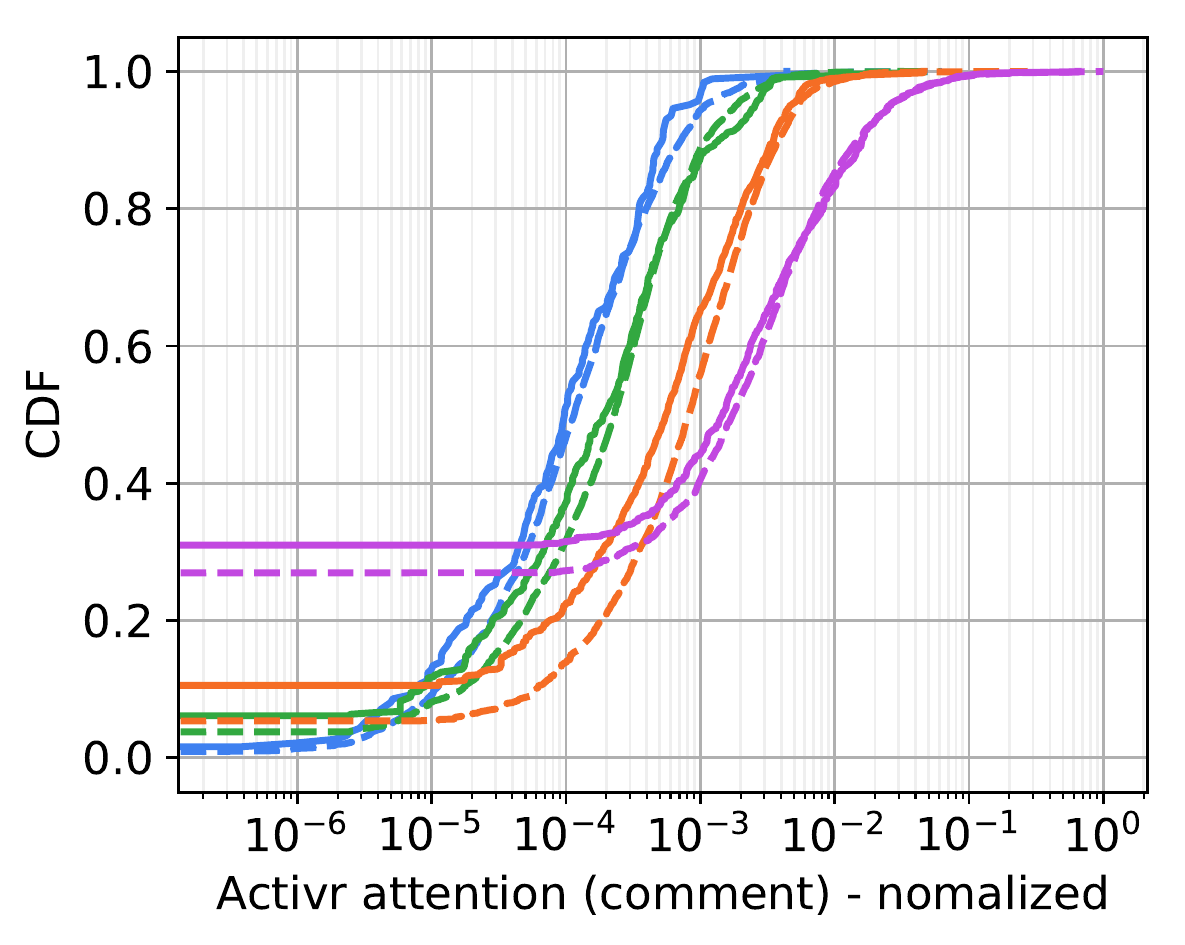}}%
 \vspace{-0.1cm}
\caption{CDF of number of comments received per-post: (a) absolute number; (b) normalized.}
\label{fig_instagram_influencer_spon_vs_nonspon_comment}
\vspace{-0.2cm}
\end{figure}

\pb{Active attention - Comments.}
Figure \ref{fig_instagram_influencer_spon_vs_nonspon_comment}(a) presents the CDF of the received comments per-post for each influencer.
We separate posts into sponsored and non-sponsored posts within the influencer timelines. 
In almost all the cases we observe that sponsored posts receive fewer comments from the users.
This suggests that these sponsored posts are of less interest than other posts.
This difference is even more significant in mega influencers, where sponsored posts gained 10 times fewer comments than their non-sponsored counterparts.
We also observe that the number of comments are ranked in order of Mega, Macro, Micro and Nano influencers with, unsurprisingly, Mega influencers getting over 40 times more comments than Nano. 

The above analysis of absolute counts may give a misleading perspective as influencers with high follower counts (\eg Mega) will obviously obtain higher comment counts. 
Hence, we normalize the comment count as a fraction of the follower count, and plot the results in Figure \ref{fig_instagram_influencer_spon_vs_nonspon_comment}(b).
Here, we see rather different trends with nano influencers gaining the most engagement.  In other words, even though popular accounts gain more comments, less popular accounts obtain engagement from a higher fraction of their follower-base. This perhaps sheds light on why nano-influencers have recently started to gain traction, with their ability to engage more targeted populations.

%

We also inspect the duration before a user adds a comment on a post. Intuitively, comments that are issued shortly after a post is created might be from more engaged users. This confirms similar results to Figure~\ref{fig_instagram_influencer_spon_vs_nonspon_comment}, with Nano influencers gaining posts most rapidly than their more popular counterparts. 
In all the cases, non-sponsored posts gain comments earlier; the most significant difference is for Mega influencers. The median comment age for a Mega influencer's non-sponsored posts is 328 \vs 366.5 for their sponsored posts.
That said, we observe a subtle difference between the influencer groups.
In the first hour $<$30\% of comments of all influencers are issued. After the first 10 hours, non-sponsored posts of Macro influencers receive $<$70\% of their total comments, but sponsored posts of mega influencers get only less than 50\%. The density of users who reacted to non-sponsored posts is larger, and this difference is more significant in mega influencers and less for nano influencers (plot not shown due to space constraints).

\begin{figure}[ht]
\vspace{-0.4cm}
    \subfloat[Passive attention (like)]{\includegraphics[width=0.49\linewidth, height=0.23\textwidth]{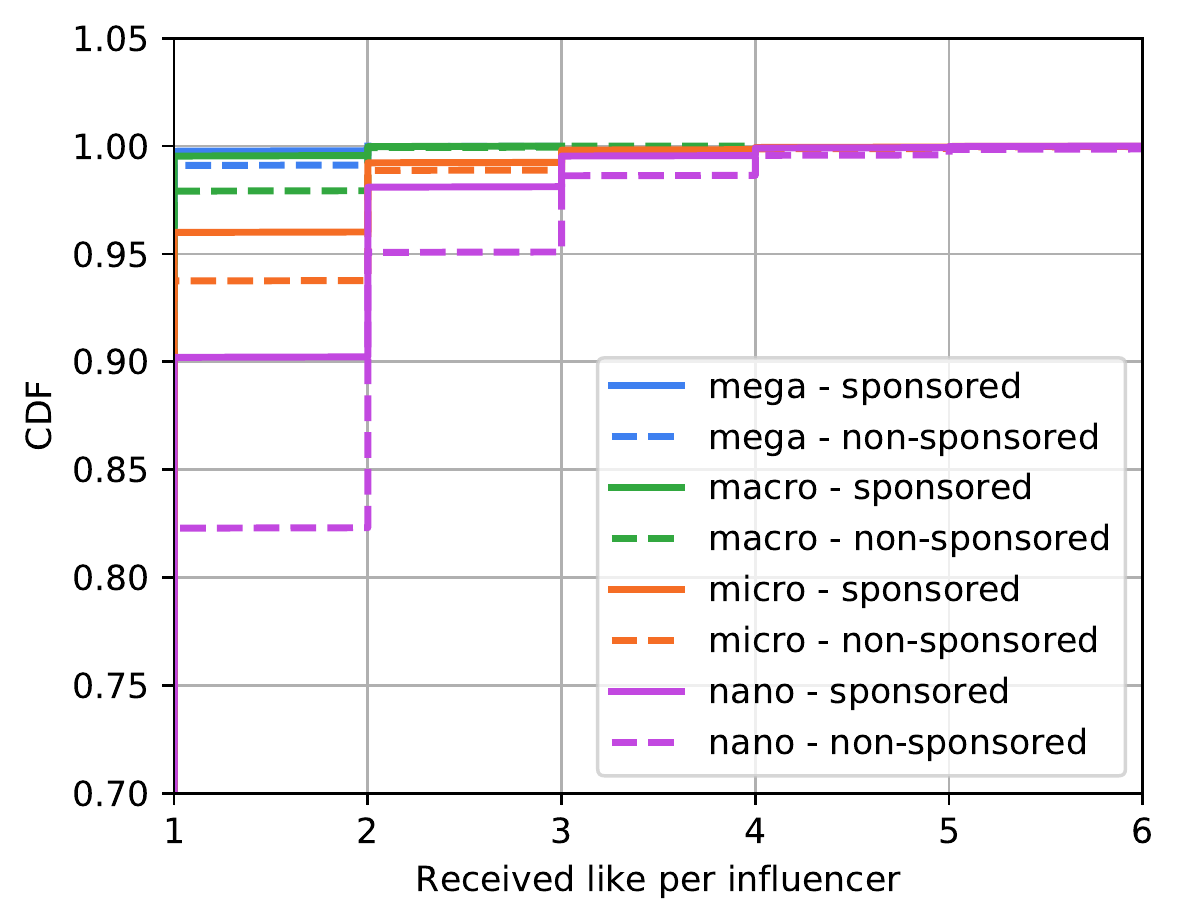}}%
  \subfloat[Active attention (comment)]{\includegraphics[width=0.49\linewidth, , height=0.23\textwidth]{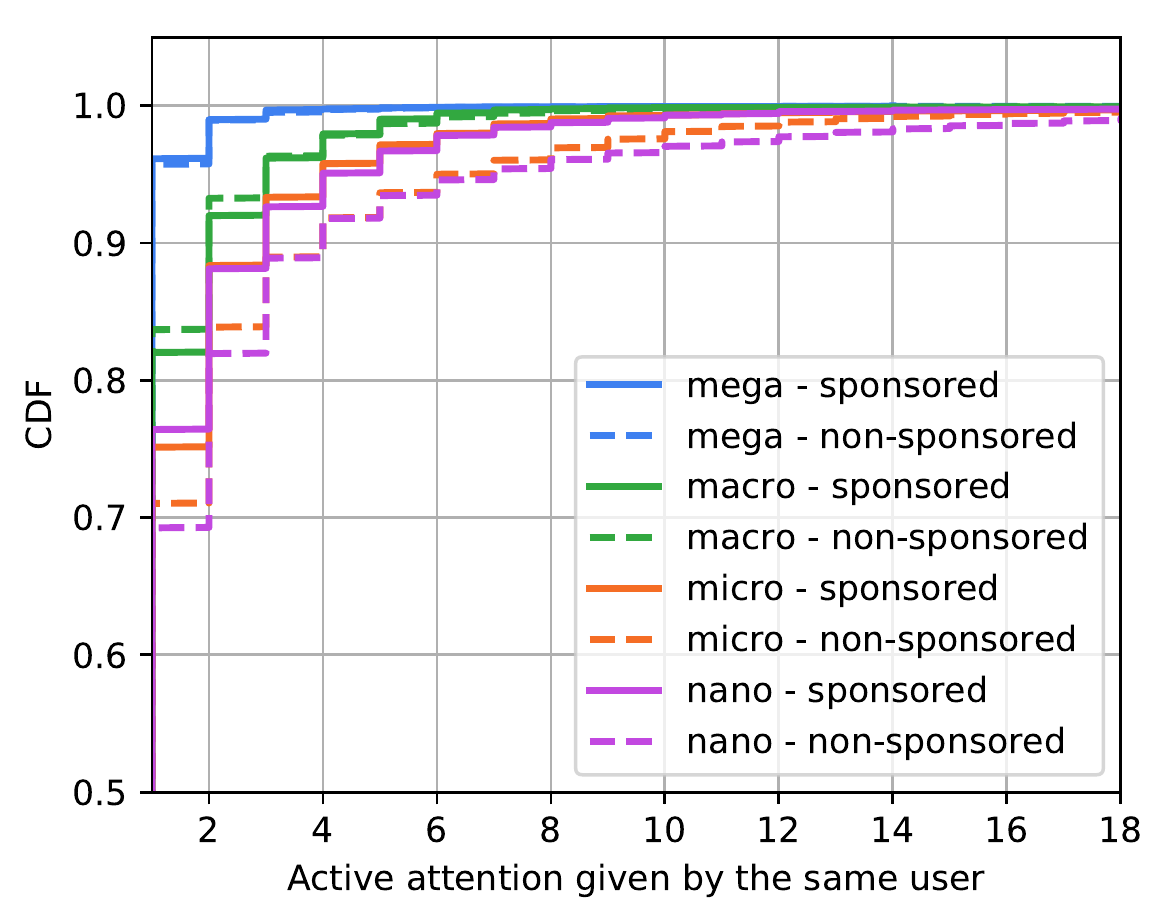}}\hfill
\caption{(a) attracted attention CDF of number of likes received per-influencer; (b) number of comments performed by each user per-influencer}
\label{fig_instagram_audience_active_by_users}
\vspace{-0.3cm}
\end{figure}

Briefly, we also note that nano influencers tend to have more consistent engagement. Whereas more popular influencers, gain comments from many different users, nano influencers tend to get comments from the same users multiple times: 30\% give more than one comment (Figure~\ref{fig_instagram_audience_active_by_users}b).

\pb{Passive attention - Likes}. 
As comments are generally the most active form of engagement, we also inspect more passive engagement via likes.
Figure~\ref{fig_instagram_audience_active_by_users}a presents the CDF of the received likes for both sponsored and non-sponsored posts among each influencer sategory. For Mega influencers, we observe 28\% more likes for non-sponsored \vs sponsored posts. This, however, is far less for the other categories (macro, micro, and nano), with an equivalent value of 6\%. In fact, we observe that sponsored posts gain marginally more likes than non-sponsored posts for nano influencers (median 56 \vs 47). 
That said, the categories exhibit broadly similar patterns to that seen in comments (with Mega gaining the most and nano gaining the fewest likes in absolute terms). Turning our attention to the normalized like count, we see that again nano influencers get more likes than the other categories. However, we do see outliers: for instance, inside nano, nearly 8\% of accounts receive a large number of likes, yet nearly no comments (all posts). To investigate this, we manually inspect these subsets of nano influencers. We find that there is a prevalence of fake profiles, apparently using bots to boost their impact.



\subsection{How often do influencers post?}
\label{influencer_posts}

\begin{figure}[htbp]
\vspace{-0.4cm}
  \subfloat[Post]{\includegraphics[width=0.45\linewidth, , height=0.21\textwidth]{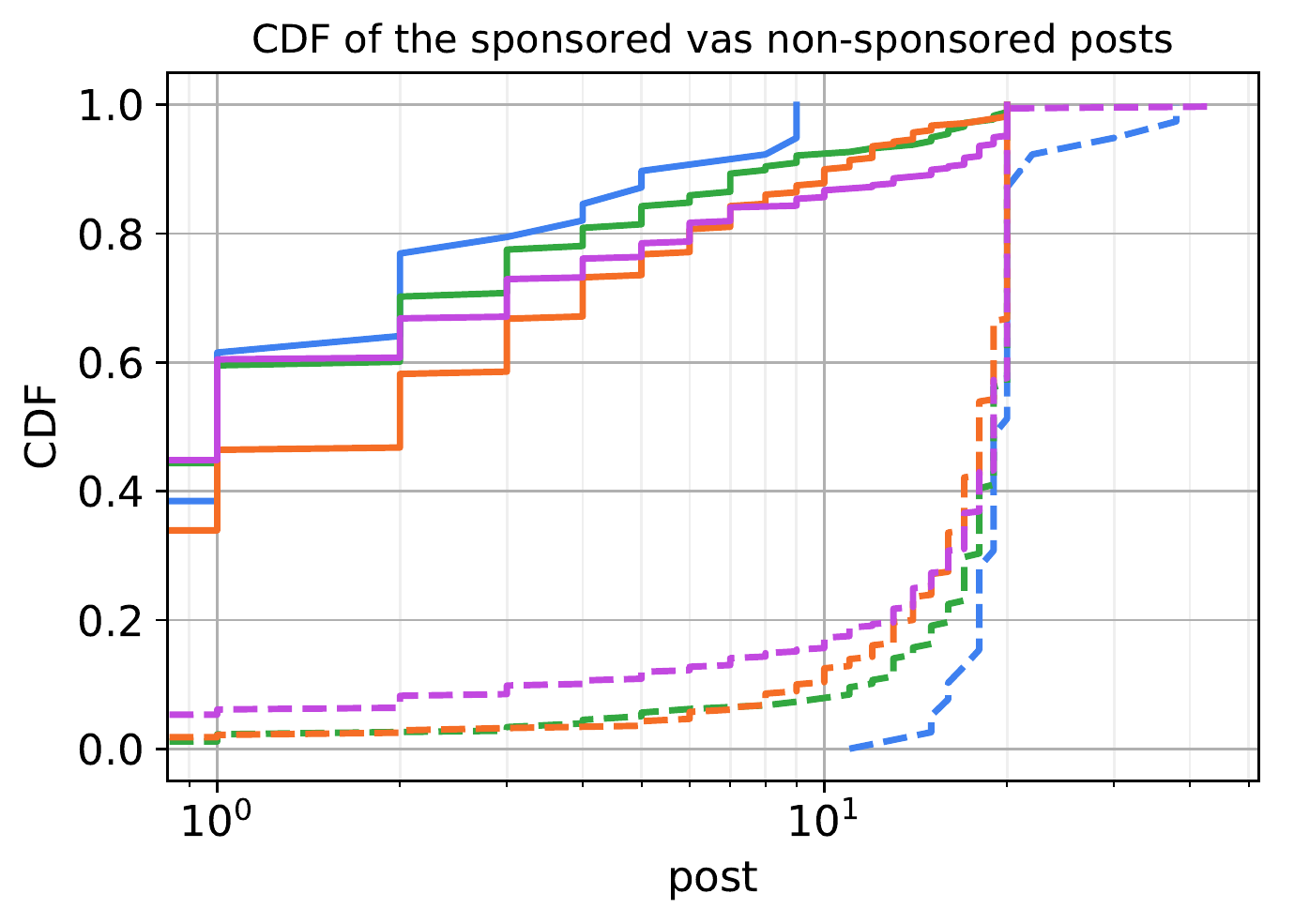}}
  \subfloat[Story]{\includegraphics[width=0.50\linewidth, height=0.21\textwidth]{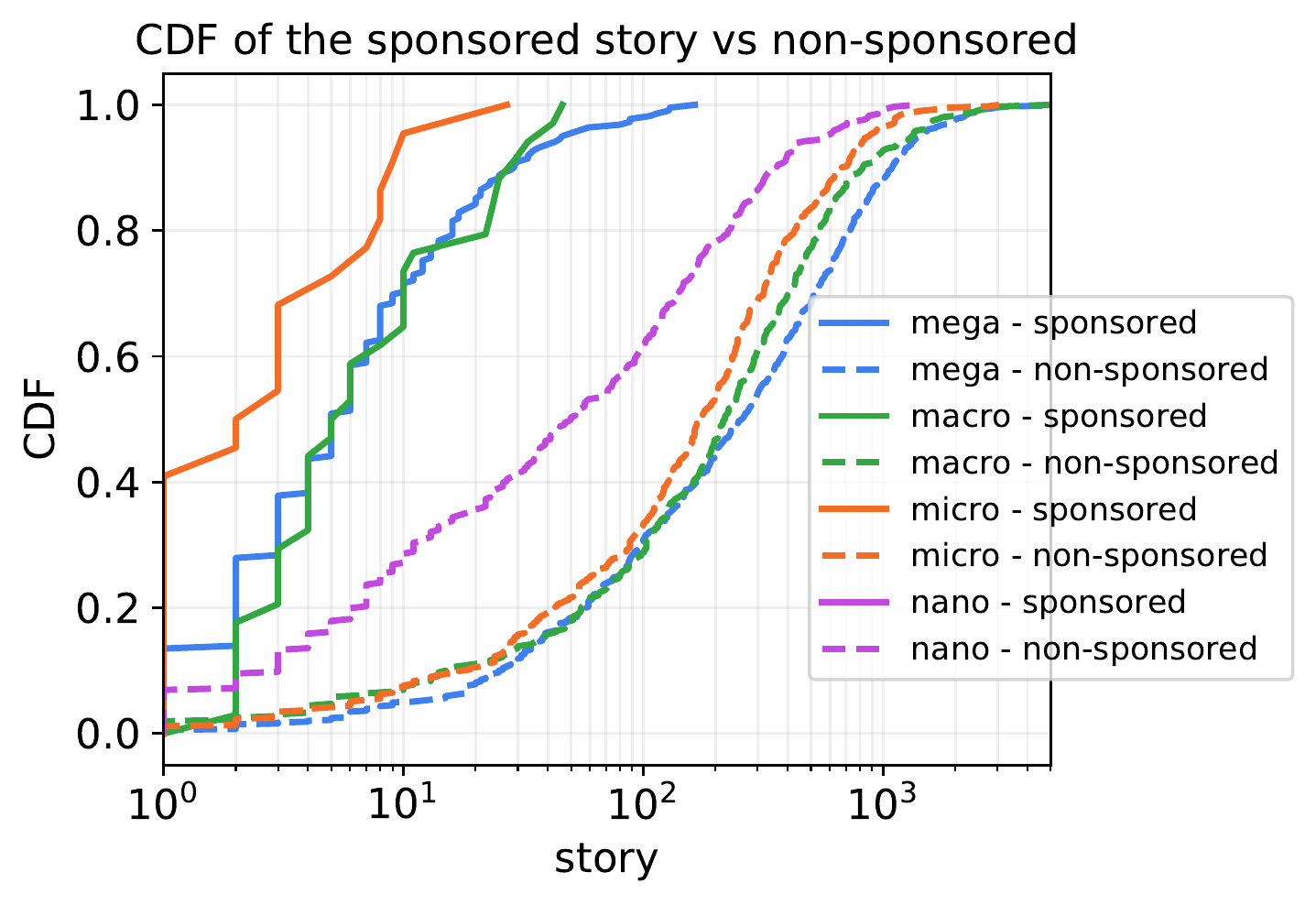}}%
  \vspace{-0.2cm}
\caption{CDF of the number of posts and stories published per influencer.}
\label{fig_instagram_influencer_spon_vs_nonspon_post}
\vspace{-0.4cm}
\end{figure}

We start by measuring the number of posts per influencer. 
Figure \ref{fig_instagram_influencer_spon_vs_nonspon_post}(a) presents a CDF plot of the number of sponsored \vs non-sponsored posts, whereas Figure~\ref{fig_instagram_influencer_spon_vs_nonspon_post}(b) repeats the same for stories.

We observe distinct distributions, with most influencers publishing more non-sponsored posts. Only 8.3\% of influencers distribute more sponsored posts compared to non-sponsored.  On average, 16\% of posts are sponsored with just 9.3\% of influencers tagging over half of their posts as sponsored. 
This is anticipated as most influencer guides recommend that users keep the percentage of sponsored posts below 60\%, to maintain audience engagement. 

Subtle differences can also be observed between the different categories of influencer. For example, where the mega influencers on average post the most sponsored posts, they actually post the least non-sponsored posts. Of course, this might also be a product of how such influencers tag their posts. We find that no sponsored story is published by any nano influencers. This contrasts to what is observed with posts from influencers within the same nano category in
Figure~\ref{fig_instagram_influencer_spon_vs_nonspon_post}(a), where over 80\% of these influencers publish $\leq$10 posts. This striking difference suggests that nano influencers are primarily using posts (rather than stories) to publish sponsored contents.
In contrast, mega influencers tend to use stories to promote sponsored contents more regularly (compared to macro and micro influencers). For example, $\leq$21\%  of Macro and $\leq$3\% of micro influencers publish more than 10 sponsored stories compared to over 30\% for mega publishers. In general, we see that influencers across mega, macro and micro category favor the use of stories to promote sponsored content compare to post, possibly because it is cheaper to advertise via stories compared to feeds~\cite{Instagra65:online}.
Another reason for using stories is the exclusivity \ie followers must stay engaged and must hurry to see the offer or discount code \etc while the story lasts.

\begin{figure}[t]
\vspace{-0.3cm}
\centerline
{\includegraphics[width=0.28\textwidth]{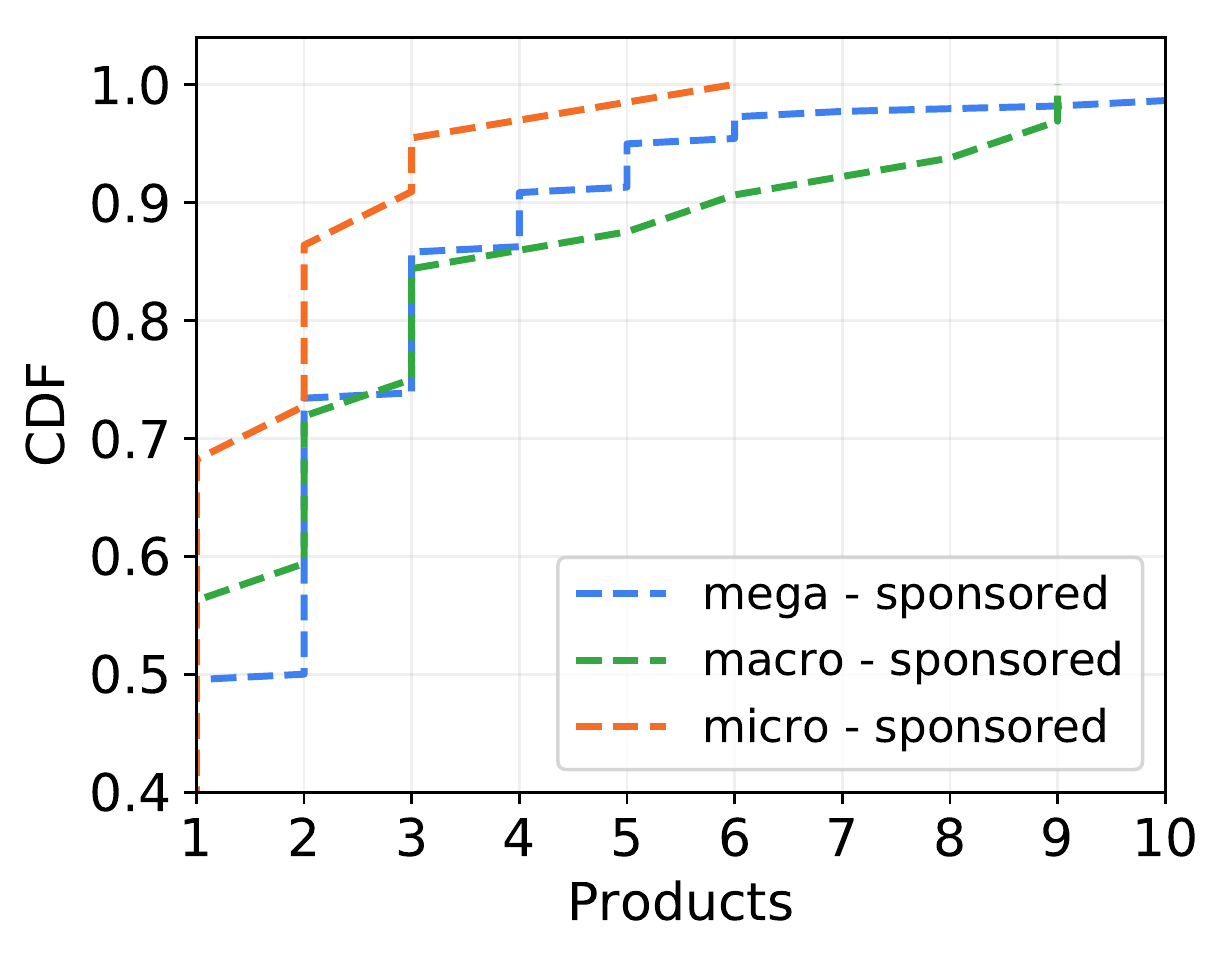}}
\vspace{-0.2cm}
\caption{CDF of number of products promoted by influencers across categories. This only covers stories because equivalent metadata is not available for posts.}
\label{figure_word_cloud_for_category_of_advertisers_scrape_further_scrape}
\vspace{-0.4cm}
\end{figure}

\subsection{What do influencers promote?}
\label{influencer_promoted_products}
Finally, we wish to inspect the types of products being promoted by influencers. 
This can be done via our Instagram stories dataset, as each sponsored item is optionally tagged with the category of the advertiser. This is taken from a control set of tags offered by Instagram. Note that this is \emph{only} available for stories, and not regular posts. 
We find that this feature is not widely used by influencers, with only 3\% stating their product. 

Figure~\ref{figure_word_cloud_for_category_of_advertisers_scrape_further_scrape} presents a CDF showing the number of products promoted by influencers across categories. Most influencers only promote a single product, particularly in the case of micro.  We observe that 50\% \emph{Mega}, 58\% \emph{Macro} and 70\% \emph{Micro} influencers promote just a single product. This suggests that influencers tend to focus on a particular product type, likely in their own specialist area.

\begin{figure}[h!]
\vspace{-0.4cm}
\centerline
{\includegraphics[width=0.45\textwidth]{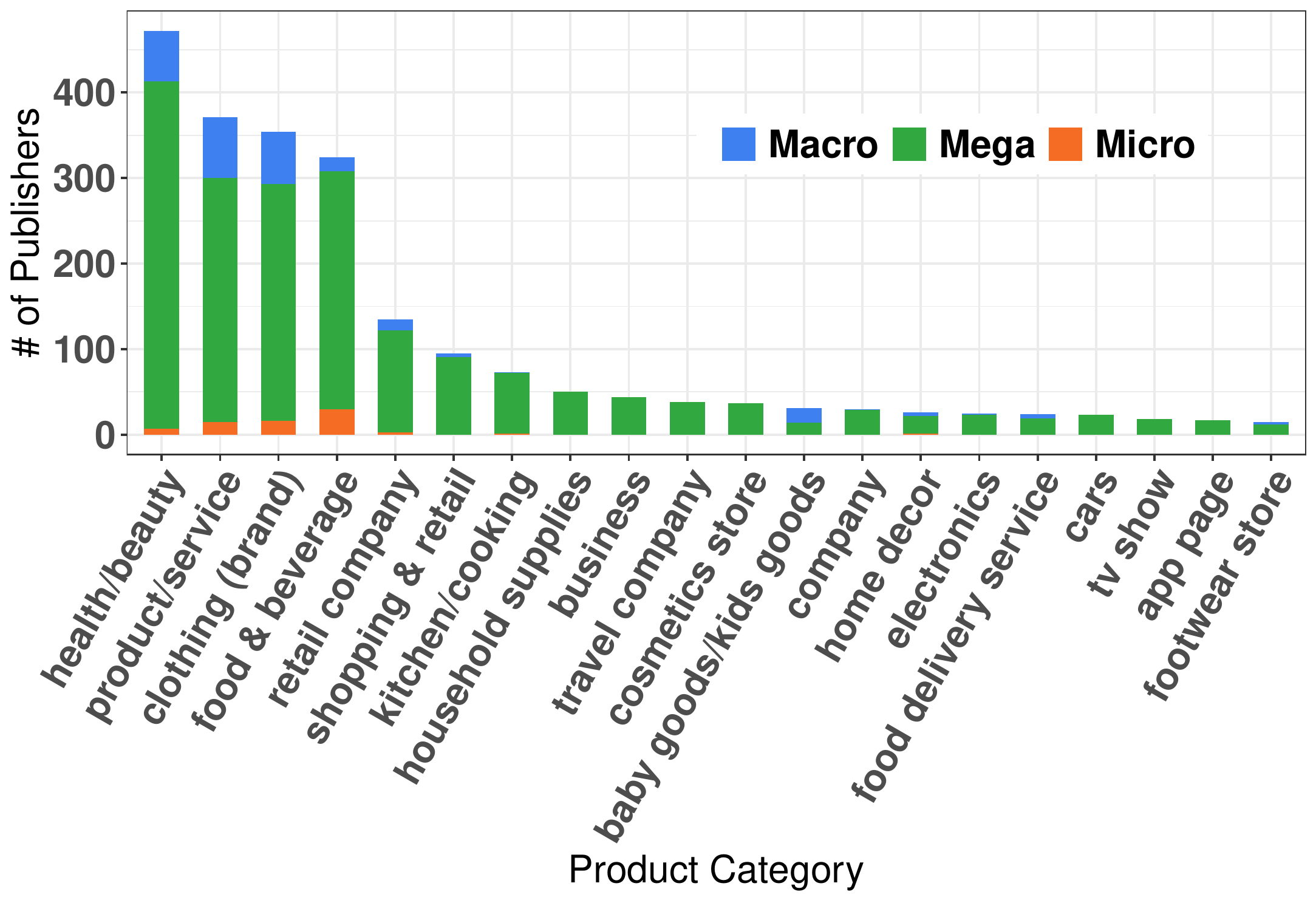}}
\vspace{-0.3cm}
\caption{Number of products promoted in stories based on their type (identified via the Instagram API).}
\label{fig_instagram_story_advertizers}

\vspace{-0.3cm}
\end{figure}

We also inspect the type of products influencers use stories to promote. This metadata is captured by the Instagram API, although it only covers stories, as posts do not contain this explicit metadata.
Figure~\ref{fig_instagram_story_advertizers} presents the top 20 product types influencers promote. The Y-axis counts the number of unique accounts promoting each type of product. We observe that most publishers tend to advertise products under \emph{Health/Beauty} (14\%), \emph{Product/Service} (11\%) and  \emph{Clothing (Brand)} (11\%).  Across products we observe that Mega influencers tend to dominate: they are the major publishers for 77\% of product types we identify. This is in sharp contrast to Macro (14\%) and for Micro (9\%). 
These findings confirm the intuition that Instagram is dominated by promotions surrounding consumables such as food, retail and beauty. These cover 43\% of all adverts.

\section{Predicting Sponsored Content}
\label{sec:predicting}

The previous results have shown that sponsored posts tend to have unique properties that differentiate them from non-sponsored content. This leads us to inspect the viability of automatically detecting (and tagging such posts). This, for example, could be used to protect users from being misled. 


\subsection{Classifier Design}
\label{subsec:classifier_design}

We take a supervised deep learning approach, as we have a substantial ground-truth (annotated) dataset of sponsored \vs non-sponsored posts.

\pb{Dataset.}
For classification, we use the post dataset described in Section~\ref{instagram_crawler}. In total, the dataset consists of nearly 7K sponsored and 27K non-sponsored posts from all groups of influencers. Note that we do not differentiate between the different popularity groups and, instead, treat all accounts equally.
As we are dealing with an imbalanced dataset, we use Random Under-Sampling to reduce the size of the non-sponsored class. Accordingly, we randomly select posts from non-sponsored class and remove them from the training dataset. While selecting the posts, we try to keep the diversity of influencer's profiles. So, we select random samples from influencers who have more published post. This process started from nano (with a higher number of post) to mega. 

Finally, we produce a dataset with 14k observations with an equal number of sponsored and non-sponsored labelled posts (Section \ref{subsection_detect_sponsored}). 
Note that all these posts are generated by users who have posted at least one sponsored post. Hence, our classification task is limited to differentiating between sponsored and non-sponsored posts from influencers (rather than a more general audience).
Based on our observations, the posts in the above dataset can be separated into three sup-populations:

\begin{enumerate}
\item \pb{Sponsored Post.} In a sponsored post, influencers normally try to directly or indirectly advertise a product. By adding sponsored metadata (Section \ref{subsection_detect_sponsored}), the post is explicitly declared as an advertisement. It may also contain related hashtags for that product and the page of that product. We note that the main company is often tagged in the post. 

\item \pb{Non-Sponsored Post.} In contrast to a Sponsored Post, this is a routine post by an influencer which does not include any direct or indirect advertisements. As a result, the photo/video, caption, and list of hashtags are not organized or written in such a way to promote anything.

\item \pb{Hidden Sponsored Post.} This sub-population is similar to the first item (Sponsored Post) except that sponsored metadata (\ie hashtags) are removed. We remind the reader that The Advertising Standards Authority (ASA)~\cite{online_asa} advises the use of such hashtags, as any paid promotions are subject to consumer protection law. Despite this, we posit that there is an incentive for influencers to avoid disclosure, as there is greater value in ``personal'' endorsements.
``\textit{Hidden Sponsored Post}",  ``\textit{Hidden Advertisement}" and "\textit{Non-declared Sponsored Post}" terms address the same meaning in this paper. 


\end{enumerate}

\begin{figure}[t]
\vspace{-0.3cm}
\centerline
{\includegraphics[width=0.48\textwidth]{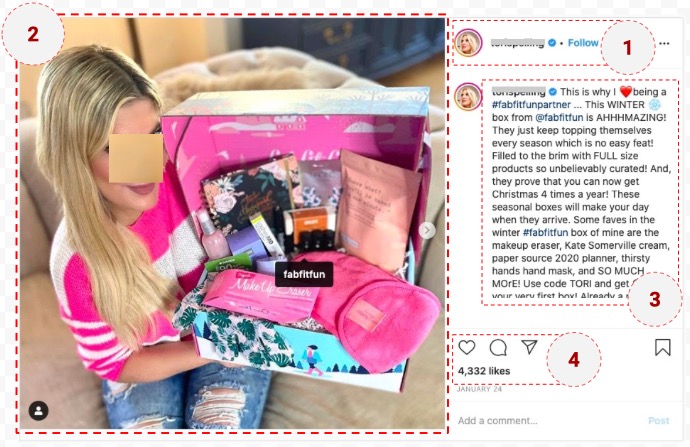}}
\vspace{-0.1cm}
\caption{An Example of non-declared sponsored post: 1) A verified Mega influencer holding a box on her hand and promoting products. 2) The product page is tagged in the photo and mentioned in the caption. 3) The caption invites us to use this product, but there is no sign of sponsored metadata. 4) The post received 4K likes and 500 comments.
}
\label{fig_hidden_post}
\vspace{-0.4cm}
\end{figure}

\pb{Feature Engineering \& Pre-processing.}
The next step is to build a set of features from the post content to train our model. Table \ref{table_feature_list} summarizes the features that are employed. We split the feature list into two main categories: post features and publisher features. Post features comprise all features that are obtained from the content of the post, \eg number of likes, the caption. Publisher features cover features extracted from the profile of the publisher. For all text-based features such as ``post caption" and ``post hashtag", we remove all punctuation marks, stopwords and convert them to lowercase characters. Words are stemmed to reduce to their root forms.

\begin{table}[h!]
\vspace{-0.2cm}
\centering
\caption{Feature Set Used for Classification.}
\vspace{-0.2cm}
\label{table_feature_list}
\resizebox{\columnwidth}{!}{%
\begin{tabular}{llll}
\multicolumn{2}{c}{\textbf{Post Features}} & \multicolumn{2}{c}{\textbf{Publisher Features}} \\ \hline
\textbf{Feature} & \multicolumn{1}{l|}{\textbf{Type}} & \textbf{Feature} & \textbf{Type} \\ \hline
post caption & \multicolumn{1}{l|}{\textit{text}} & number of follower & \textit{numeric} \\
post hashtag & \multicolumn{1}{l|}{\textit{text}} & number of followee & \textit{numeric} \\
number of likes & \multicolumn{1}{l|}{\textit{numeric}} & length of biography & \textit{numeric} \\
number of comments & \multicolumn{1}{l|}{\textit{numeric}} & profile biography & \textit{text} \\
length of caption & \multicolumn{1}{l|}{\textit{numeric}} & is verified & \textit{numeric} \\
number of hashtags & \multicolumn{1}{l|}{\textit{numeric}} & external URL exist & \textit{numeric} \\
number of mentions & \multicolumn{1}{l|}{\textit{numeric}} &  &  \\
number of tagged users & \multicolumn{1}{l|}{\textit{numeric}} &  & \textit{} \\ \hline
\end{tabular}
}
\vspace{-0.2cm}
\end{table}

\begin{figure}[h!]
\vspace{-0.2cm}
\centerline
{\includegraphics[width=0.5\textwidth]{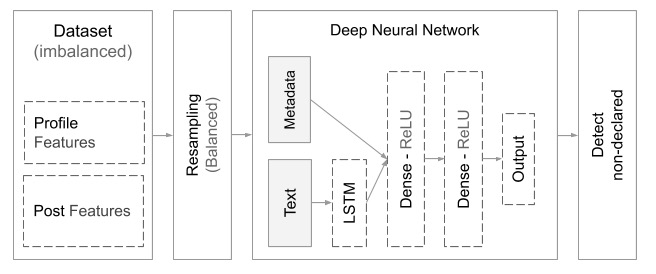}}
\vspace{-0.3cm}
\caption{General architecture of sponsored content detection (Contextual LSTM model).}
\label{fig_dnn_architecture}
\vspace{-0.2cm}
\end{figure}

\pb{Model Architecture \& Performance.} 
Next, we propose a Contextual LSTM Neural Network architecture (presented in Figure \ref{fig_dnn_architecture}) and we compare the result with Random Forest as a classic machine learning based Classifier.
Post metadata can enrich the information available for classification. So, to build the model, we combine texts with other metadata features (Table \ref{table_feature_list}). First, we tokenize text metrics (\eg{}caption and hashtag) using Keras Tokenizer Class \cite{chollet2015keras} and then, the result is fed to the LSTM layer which outputs a 64-dimension vector. Then, we attach numerical metadata (post and profile features) to this vector and pass it through 2 ReLU activated layers of size 128 and 64. Finally, it connects to an output layer that predicts the label.

We use TensorFlow, Keras, and Scikit-learn libraries, and we run them on Google CoLab. Also, we use a random split of 80\% (training set) and 20\% (test set) and to avoid overfitting, we use 10-fold Cross-validation. For each model, we compute the accuracy, precision, recall and F1 in Table~\ref{table_model_performance}.
For our labelled data, we observe high performance, with positive results across both classifiers. Our Contextual LSTM Classifier enhances the performance by nearly 5\% (89\% of accuracy).


\begin{table}[t!]
\centering
\caption{Classifier Performance}
\label{table_model_performance}
\resizebox{\columnwidth}{!}{%
\begin{tabular}{l|rrrr}
\textbf{Model} & \textbf{Accuracy} & \textbf{Precision} & \textbf{Recall} & \textbf{F1} \\ \hline
\multicolumn{1}{l|}{Random Forest} & 0.84 & 0.83 & 0.84 & 0.83 \\
\multicolumn{1}{l|}{Contextual LSTM} & 0.89 & 0.88 & 0.87 & 0.89
\end{tabular}
}
\vspace{-0.3cm}
\end{table}

\subsection{Results}\label{hidden_sponsored}
We now evaluate our classifier across the entire dataset to quantify the prevalence of hidden sponsorship.

\pb{Detection.} As we believe some influencers do not declare sponsored posts with defined metadata (\S\ref{subsection_detect_sponsored}), we use the DNN trained model on the non-sponsored posts which are not used in the training. This part includes nearly 20K posts from all 4 groups of influencers (mega to nano). 

\begin{figure}[h!]
\vspace{-0.3cm}
\centerline
{\includegraphics[width=0.45\textwidth]{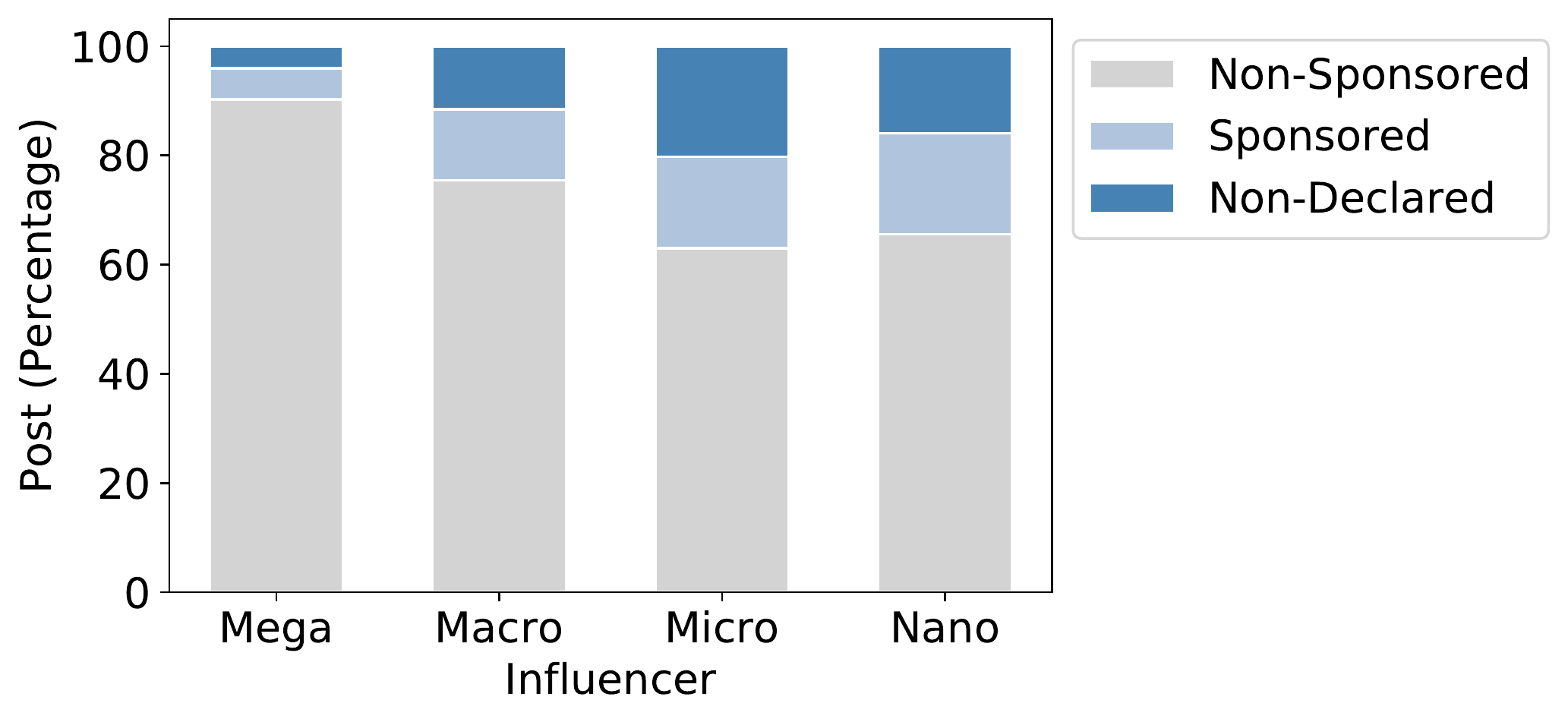}}
\vspace{-0.2cm}
\caption{Overall view of different post distributed by influencers.}
\label{figure_hidden_post_percentage}
\vspace{-0.3cm}
\end{figure}

\pb{Results.}
Our model detects that 17.7\% sponsored posts are published without metadata (or not declared as sponsored) which is approximately 3.8K posts through all influencers.
Figure \ref{figure_hidden_post_percentage} demonstrates the portion of \textit{non-sponsored}, \textit{sponsored} and \textit{non-declared sponsored} post across each group. 
We see a noticeable set of hidden sponsored posts, particularly for Micro and Nano influencers. Nano with 18\%, micro with 21\%, and macro with 11\% of shared posts, advertise almost 4 times more than Mega. In detail, by considering clusters of influencers we observe:

\begin{itemize}
\item \pb{Mega.} Mega is the smallest group in terms of having hidden advertisements. Nearly 29\% of Mega influencers publish one non-declared post. In detail, out of 3K non-sponsored posts, 4.3\% are detected as hidden advertisement entities. 

\item \pb{Macro.} 57\% of Macro influencers have one hidden sponsored post. In detail, 13\% of 3.6K non-sponsored posts are recognised as hidden advertisements.

\item \pb{Micro.} Micro is the largest group in terms of publishing non-declared advertisements. In general, 78\% of Micro influencers publish one sponsored post. In detail, 24\% of the 5K non-sponsored content is detected as hidden advertisements. 

\item \pb{Nano.} More than 70\% of nano influencers publish one non-declared advertisement. In detail, within 9.7K non-sponsored posts, 20\% are detected as hidden advertisements.

\end{itemize}

\pb{Validation.} To validate the above results, we manually validate 50\% of the detected posts to confirm they are hidden sponsored posts. 
Across the identified content, we witness in 81\% of posts, influencers are confirmed as promoting products, 11\% are incorrect, and in 8\% we are not able to verify they are commercial or not. This confirms that our model \emph{is} effective at identifying influencer posts.

\subsection{Text Feature Analysis}
\label{subsec:feature_analysis}
Finally, we manually explore which textual features (Table \ref{table_feature_list}) are most prominent in our prediction task. We do this to gain an understanding of what characteristics are most common specifically for (hidden) sponsored posts. 
Thus, we manually check 50\% of the predicted posts and for each metric, we inspect the unique distinctions between sponsored and non-sponsored posts. 
This allows us to compile a list of differences and then count how often these difference emerge. We summarise our findings as follows:


\begin{enumerate}

\item \pb{Post Caption} 
holds the most valuable information of the post in the shape of the text  (Fig \ref{fig_hidden_post}). Often the text includes details of products, promotions, names, \etc and encourages users to perform an action. An action could be to buy something, follow a page, join a competition, install an application, use a specified hashtag, \etc 
We also notice some peculiar way of writing the text or using keywords as follows. 
If X represents a \textit{product} or \textit{producer}: 
\one~the caption includes X as raw text in 94\% of the sponsored posts. 
\two~Also X is mentioned (followed by @ sign and becomes blue) in 91\% of the posts.
\three~The caption could contain particular sentences such as ``\textit{thank you X}", ``\textit{many thanks to X}", ``\textit{X from this page}", ``\textit{my top choice is X}", ``\textit{go and follow X}", \etc. This happens in 78\% of the sponsored posts.
\four~We observe call-to-action keywords such as ``\textit{link in bio}", ``\textit{download it}", ``\textit{watch my story}", ``\textit{Use discount code}" and ``\textit{comment/like to win}", \etc in 53\% of the posts.

\item \pb{Post Hashtag} 
is the next leading text feature which generally includes valuable information about the product and its producer. Influencers may use one or more hashtags for the corresponding product (Figure \ref{fig_hidden_post}). For example, if \textit{\#outfit} is the main hashtag representing the product, we also observe \textit{\#outfitday}, \textit{\#oufit}, \textit{\#bestoutfit}, \textit{\#trendyoutfit}, \textit{\#outfitstyle}, \etc 
In 97\% of the sponsored posts, the name of the product or producer (or both) is listed as a hashtag. 
To highlight some hashtags, influencers may put them in the text while describing the products, or create a separate list after the text (or in some cases in the first comment by the influencer). 
Hashtag count also helps differentiate sponsored content. Sponsored posts one{} to get hashtags trending, or \two{} to get more visibility in Instagram explorer, or \three{} to be easily found in search results, regularly have longer hashtag length: \

\item \pb{Profile Biography} 
is another feature that improves the accuracy of our classifier. In general, influencers usually put relevant information in their profile biography (which could be temporary): 
\one~In 63\% of the biographies, information such as ``sponsor info", ``campaign details" or ``promotion code" as raw text exist.
\two~In 54\% of the profiles, influencers put sponsor ``hashtag(s)" (which becomes blue) or mention ``product/producer page" (followed by @ sign). 
\three~In 34\% of the profiles, there is a call-to-action phrase with special keywords such as "follow", ``buy", ``sale", ``watch", ``join", ``check out", ``promotion", ``more info'' \etc
\four~Profiles may include an External URL (in 21\% of the profiles) or YouTube link (in 11\% of the profiles), which redirects users to the main product webpage, producer website, or full youtube review video. 
Regularly influencers use shortener tools to monitor the number of people clicking on it. 
Biography length is also helpful as influencers who do promotion usually have \one{} promotion codes, \two{} sponsor contact detail, and \three{} product-related hashtags on their biography (longer biography).

\end{enumerate}

\section{Conclusion \& Future Work}
This paper has performed the first large-scale analysis of influencer behaviour on Instagram. Rather than solely discovering ``celebrities'', our methodology has exposed a large array of influencer types, including highly targeted nano influencers. We have further dissected the differing behaviours of these influencers, to find that even though Mega influencers garner significant attention, nano influencers are more successful in engaging their audiences with a larger fraction of followers interacting with their posts. 
We have further trained a DNN model to classify posts as sponsored, allowing us to identify seemingly sponsored posts that are not properly declared. As influencer income is taxable, we note that this may have financial ramifications that go beyond the issues of deceiving consumers. 



This is just the first step in a wider research agenda. We have several specific lines of future work. 
First, we wish to expand our analysis across multiple platforms (\eg Tik Tok, YouTube) and to gain a deeper understanding of the strategies employed by influencers. 
Second, we wish to revisit our classifier to improve performance. Although our current implementation obtains good results, we manually found 11\% misclassifications. 
Consequently, we wish to build on our work to devise classifiers that can automatically identify influencers. This is particularly important for identifying ``ghost'' influencers who do not explicitly identify their adverts. We posit that this could also be used to support social media giants in identifying misbehaviour within the advertisement space.
Whereas in this paper, we have focused on influencers themselves, we are also curious to better understand the dynamics of \emph{audiences} as well as the \emph{companies} that use influencers.


\bibliographystyle{unsrt}
\bibliography{bib}

\begin{thebibliography}{10}

\bibitem{CASALO2018}
Luis~V. Casaló, Carlos Flavián, and Sergio Ibáñez-Sánchez.
\newblock Influencers on instagram: Antecedents and consequences of opinion
  leadership.
\newblock {\em Journal of Business Research}, 2018.

\bibitem{asa_guidelines}
ASA.
\newblock An influencer’s guide to making clear that ads are ads, 2018.

\bibitem{rogers2010diffusion}
Everett~M Rogers.
\newblock {\em Diffusion of innovations}.
\newblock Simon and Schuster, 2010.

\bibitem{SINGH201987}
Amandeep Singh, Malka~N. Halgamuge, and Beulah Moses.
\newblock 5 - an analysis of demographic and behavior trends using social
  media: Facebook, twitter, and instagram.
\newblock In Nilanjan Dey, Samarjeet Borah, Rosalina Babo, and Amira~S. Ashour,
  editors, {\em Social Network Analytics}, pages 87 -- 108. Academic Press,
  2019.

\bibitem{hennig2004electronic}
Thorsten Hennig-Thurau, Kevin~P Gwinner, Gianfranco Walsh, and Dwayne~D
  Gremler.
\newblock Electronic word-of-mouth via consumer-opinion platforms: what
  motivates consumers to articulate themselves on the internet?
\newblock {\em Journal of interactive marketing}, 18(1):38--52, 2004.

\bibitem{brown2008influencer}
Duncan Brown and Nick Hayes.
\newblock {\em Influencer marketing}.
\newblock Routledge, 2008.

\bibitem{biaudet2017influencer}
Sofie Biaudet.
\newblock Influencer marketing as a marketing tool: The process of creating an
  influencer marketing campaign on instagram.
\newblock 2017.

\bibitem{ewers2017sponsored}
NL~Ewers.
\newblock \# sponsored--influencer marketing on instagram: An analysis of the
  effects of sponsorship disclosure, product placement, type of influencer and
  their interplay on consumer responses.
\newblock Master's thesis, University of Twente, 2017.

\bibitem{vaibhavi_nandagiri_2018_1207039}
Vaibhavi Nandagiri and Leena Philip.
\newblock {Impact of Influencers from Instagram and YouTube on their
  Followers}.
\newblock {\em {International Journal of Multidisciplinary Research and Modern
  Education}}, 2018.

\bibitem{glucksman2017rise}
Morgan Glucksman.
\newblock The rise of social media influencer marketing on lifestyle branding:
  A case study of lucie fink.
\newblock {\em Elon Journal of Undergraduate Research in Communications},
  8(2):77--87, 2017.

\bibitem{freberg2011social}
Karen Freberg, Kristin Graham, Karen McGaughey, and Laura~A Freberg.
\newblock Who are the social media influencers? a study of public perceptions
  of personality.
\newblock {\em Public Relations Review}, 37(1):90--92, 2011.

\bibitem{sammis2015influencer}
Kristy Sammis, Cat Lincoln, and Stefania Pomponi.
\newblock {\em Influencer marketing for dummies}.
\newblock John Wiley \& Sons, 2015.

\bibitem{lahuerta2016looking}
Eva Lahuerta-Otero and Rebeca Cordero-Guti{\'e}rrez.
\newblock Looking for the perfect tweet. the use of data mining techniques to
  find influencers on twitter.
\newblock {\em Computers in Human Behavior}, 64:575--583, 2016.

\bibitem{kim2017social}
Seungbae Kim, Jinyoung Han, Seunghyun Yoo, and Mario Gerla.
\newblock How are social influencers connected in instagram?
\newblock In {\em International Conference on Social Informatics}, pages
  257--264. Springer, 2017.

\bibitem{FTC:online}
US~Federal~Trade Commission.
\newblock Us federal trade commission.
  https://www.ftc.gov/news-events/press-releases/2017/09/csgo-lotto-owners-settle-ftcs-first-ever-complaint-against.

\bibitem{InstagramAPI_online}
Instagram.
\newblock Official api graph instagram.
\newblock \url{https://developers.facebook.com/docs/instagram-api}, September
  2019.

\bibitem{InstagramAPIHashtag:online}
Instagram.
\newblock Instagram hashtag search.
\newblock
  \url{https://developers.facebook.com/docs/instagram-api/guides/hashtag-search},
  September 2019.

\bibitem{cha2010measuring}
Meeyoung Cha, Hamed Haddadi, Fabricio Benevenuto, and Krishna~P Gummadi.
\newblock Measuring user influence in twitter: The million follower fallacy.
\newblock In {\em fourth international AAAI conference on weblogs and social
  media}, 2010.

\bibitem{InstagramRqe:online}
Instagram.
\newblock What are the requirements to apply for a verified badge?
\newblock \url{https://help.instagram.com/312685272613322}, September 2019.

\bibitem{Instagra65:online}
Json Hjh.
\newblock Instagram stories vs feed ads. which is more effective at driving
  traffic?
\newblock
  \url{https://www.agorapulse.com/social-media-lab/instagram-stories-ads},
  February 2018.

\bibitem{online_asa}
The Advertising Standards Authority~Ltd. (trading~as ASA).
\newblock The advertising standards authority. https://www.asa.org.uk.

\bibitem{chollet2015keras}
Fran\c{c}ois Chollet et~al.
\newblock Keras.
\newblock \url{https://keras.io}, 2015.

\end{thebibliography}

\end{document}